\documentclass{lpor}
\usepackage[english]{babel}
\usepackage{makecell}
\usepackage{multirow}

\category{Original Paper} %

\title{Observation of SQUID-like behavior in fiber laser with intra-cavity epsilon-near-zero effect}

\author[J. Wu]{Jiaye Wu\inst{1,5,\dag}}
\author[X.Liu]{Xuanyi Liu\inst{2,\dag}}
\author[B. A. Malomed]{Boris A. Malomed\inst{3,4}}
\author[K. C. Chang]{Kuan-Chang Chang\inst{1}}
\author[M. Zhao]{Minghe Zhao\inst{1}}
\author[K. Qi]{Kang Qi\inst{1}}
\author[Y. Sha]{Yanhua Sha\inst{1}}
\author[Z. T. Xie]{Ze Tao Xie\inst{1}}
\author[M. Clementi]{Marco Clementi\inst{5}}
\author[C.-S. Br\`{e}s]{Camille-Sophie Br\`{e}s\inst{5}}
\author[S. Zhang]{Shengdong Zhang\inst{1}}
\author[H. Y. Fu]{H. Y. Fu\inst{2}\footnote{Corresponding author\quad E-mail:~\textsf{hyfu@sz.tsinghua.edu.cn}}}
\author[Q. Li]{Qian Li\inst{1}\footnote{Corresponding author\quad E-mail:~\textsf{liqian@pkusz.edu.cn}}}

\address[0]{\dag These authors contribute equally: Jiaye Wu, Xuanyi Liu.}
\address[1]{School of Electronic and Computer Engineering, Peking University, Shenzhen 518055, China.}
\address[2]{Tsinghua Shenzhen International Graduate School, Tsinghua University, Shenzhen 518055, China.}
\address[3]{Department of Physical Electronics, School of Electrical Engineering, Faculty of Engineering, and Center for Light-Matter Interaction, Tel Aviv University, Tel Aviv 69978, Israel.}
\address[4]{Instituto de Alta Investigaci\'{o}n, Universidad de Tarapac\'{a}, Casilla 7D, Arica, Chile.}
\address[5]{ \'{E}cole Polytechnique F\'{e}d\'{e}rale de Lausanne (EPFL), Photonic Systems Laboratory (PHOSL), STI-IEM, Station 11, Lausanne CH-1015, Switzerland.}

\shortauthors{J. Wu, X. Liu, et al.}

\begin{abstract}
Establishing relations between fundamental effects in far-flung areas of physics is a subject of great interest in the current research. We here report realization of a novel photonic system akin to the radio-frequency superconducting quantum interference device (RF-SQUID), in a fiber laser cavity with epsilon-near-zero (ENZ) nanolayers as intra-cavity components. Emulating the RF-SQUID scheme, the photonic counterpart of the supercurrent, represented by the optical wave, circulates in the cavity, passing through effective optical potential barriers. Different ENZ wavelengths translate into distinct spectral outputs through the variation of cavity resonances, emulating the situation with a frequency-varying tank circuit in the RF-SQUID. Due to the presence of the ENZ element, the optical potential barrier is far lower for selected frequency components, granting them advantage in the gain-resource competition. The findings reported in this work provide a deeper insight into the ultrafast ENZ photonics, revealing a new path towards the design of nanophotonic on-chip devices with various operational functions, and offer a new approach to study superconducting and quantum-mechanical systems.
\end{abstract}
\shortabstract
\begin{document}
\maketitle

\section{Introduction}
The past two decades have seen the rapid development of epsilon-near-zero (ENZ) linear and nonlinear photonics \cite{Liberal2017, Reshef2019, Kinsey2019, Wu2021a}. Combining ultrahigh intensity and ultrashort duration of pulses with the near-zero permittivity or refractive index and large optical nonlinearities \cite{Ciattoni2016, Alam2016, Alam2018, Suresh2021}, unconventional light-matter interaction and pulse shaping \cite{Wu2019, Wu2020, Xu2020, Wu2021b}, frequency translation \cite{Khurgin2020, Zhou2020}, generation of the second-, third-, and higher harmonics \cite{Capretti2015, Capretti20151, Yang2019, Rodriguez-Sune2020, Wu2021, Tian2021}, as well as the terahertz generation \cite{Jia2021}, have been demonstrated, giving rise to nanophotonic applications, such as optical switching \cite{Guo2017, Bohn2021}, electro-optical modulation \cite{Wood2018}, photonic memory \cite{Parra2019}, etc.

The currently available studies of ENZ in optics are typically conducted in conservative settings (rather than in laser cavities), with light passing the sample a finite number of times, \textit{viz}., once (transmission), twice (reflection, disregarding internal reflections), or several times (in multilayers) \cite{Rodriguez-Sune2020, Jia2021, Yang2019}. In such cases, ENZ media can be regarded as electromagnetic (EM) ideal fluids \cite{Liberal2020}, relating ENZ photonics to fluid dynamics. However, ENZ materials can exhibit an altogether different intra-cavity phenomenology in laser cavities, involving dissipative effects, an unlimited number of passes (roundtrips), and light-matter interactions.

A completely different area of physics is based on the concept and applications of superconductivity (SC) \cite{Schafroth1954, Matthias1963,Drozdov2019}. A spectacular recent achievement in this field is the discovery of SC at 250 kelvins in lanthanum hydride under high pressure \cite{Drozdov2019}. Much earlier, a very important ramification of studies of the SC dynamics was initiated by the prediction of the Josephson effect in junctions formed by two bulk SCs separated by a narrow layer of a dielectric material \cite{Josephson1962, Josephson1974}. The physics of Josephson junctions (JJs) has itself grown into a vast field of experimental and theoretical studies \cite{Barone1982,Ustinov2015}. In particular, an SC loop interrupted by one or two weak links, in the form of JJs, is the basis of ``superconducting quantum interference devices'' (SQUIDs), which provide the most sensitive tool for measuring weak magnetic fields (for instance, in the studies of biomagnetism \cite{Sternickel2006}). The weak link represents an effective potential barrier for the charge carriers in the superconductors: tunneling across this barrier results in quantum interference at the output of the SQUID. We further remark that a (small) imaginary part of this complex potential could be used to describe any excess ohmic loss inside the system. If the loop contains a single weak-link JJ, it is called a radio-frequency (RF-) SQUID.

Of course, the nature of the Josephson effect is purely quantum-mechanical. However, it is a macroscopic quantum phenomenon, hence its observation (in particular, in the RF-SQUID configuration) and the relevant models (based on the commonly known classical sine-Gordon equation for the phase difference of the wave function of Cooper pairs across the JJ \cite{Josephson1962, Josephson1974, Barone1982,Ustinov2015}) seem as quasi-classical dynamical regimes. This fact suggests a possibility to look for similar phenomenology in classical-wave settings. In this vein, in photonics it is intuitively attractive to consider ENZ and conventional optical media as ones emulating SCs and dielectric materials, respectively \cite{Rodriguez-Fortuno2014}. Specifically, in the former case the complex refractive index -- which results in slowing down, bending, and attenuation of light -- represents the counterpart of the complex electric impedance of the SQUID operating in the RF regime. In particular, the imaginary part of the effective index represents the photonic counterpart of the electric inductance in the RF-SQUID scheme. The fiber-laser--SQUID similarity may be further extended for media combining symmetrically placed gain and loss elements, in the framework of parity-time symmetric optical \cite{ElGanainy2007} and electric \cite{Ramezani2012} systems.

In superconductivity and optics alike, the essence of the SQUID-like behavior is that, when a certain physical parameter related to the SC-dielectric junction (or, in the case of its optical counterpart, the ENZ-non-ENZ junction) changes, the oscillation frequency of the tank circuit (TC, or the mode-locking frequency in the laser cavity, in terms of the optical counterpart) shifts. Inspired by this principle, in this work we hypothesize a mode-locking wavelength shift due to the quasi-SC role of the ENZ material, which is represented by its complex refractive index. We propose a phenomenological model of intra-cavity mode reselection, which assumes that, in the ENZ setup, certain frequency components pass a far lower optical potential barrier, thus acquiring an advantage in the gain-resource competition. The overall operation scheme resembles the one implemented in the RF-SQUID by means of the frequency variation in the TC. In this context, it is relevant to stress that, while SQUID is a tool for the precise detection of weak magnetic fields, the fiber-laser scheme, experimentally elaborated in this work, offers a similarly operating system for the identification of ENZ wavelength, $\lambda_{\mathrm{ENZ}}$.

Thus, we design a corresponding experimental setup and demonstrate the novel phenomenon similar to the quantum-mechanical operation of the SQUID, \textit{viz}., RF-SQUID-like behavior in a fiber laser with intra-cavity ENZ effects controlled by a variable resonance frequency. When the EM wave passes the optical potential barrier induced by the ENZ material, different ENZ wavelengths $\lambda_{\mathrm{ENZ}}$ lead to distinct spectral outputs via wavelength reselection and spectrum redistribution; while for the setting without the ENZ element a CW output is obtained, instead of the mode-locking (ML) regime, due to the polarization-evolution mismatch and strong absorption. This fact enables us to identify $\lambda_{\mathrm{ENZ}}$ directly within the cavity, by means of comparison of different intra-cavity specimens providing the ENZ effect. The results reported below provide deeper insight into ultrafast ENZ photonics and unveil a pathway towards the design of novel nanophotonic on-chip optical devices with various functionalities. The results also suggest new possibilities for the study of the SQUID-like behavior at optical frequencies, several orders of magnitude beyond the operation speed of a traditional RF-SQUID.

\section{The RF-SQUID and SQUID-like Behavior}

\begin{figure}[th]
	\centering
	\includegraphics[width=1\linewidth]{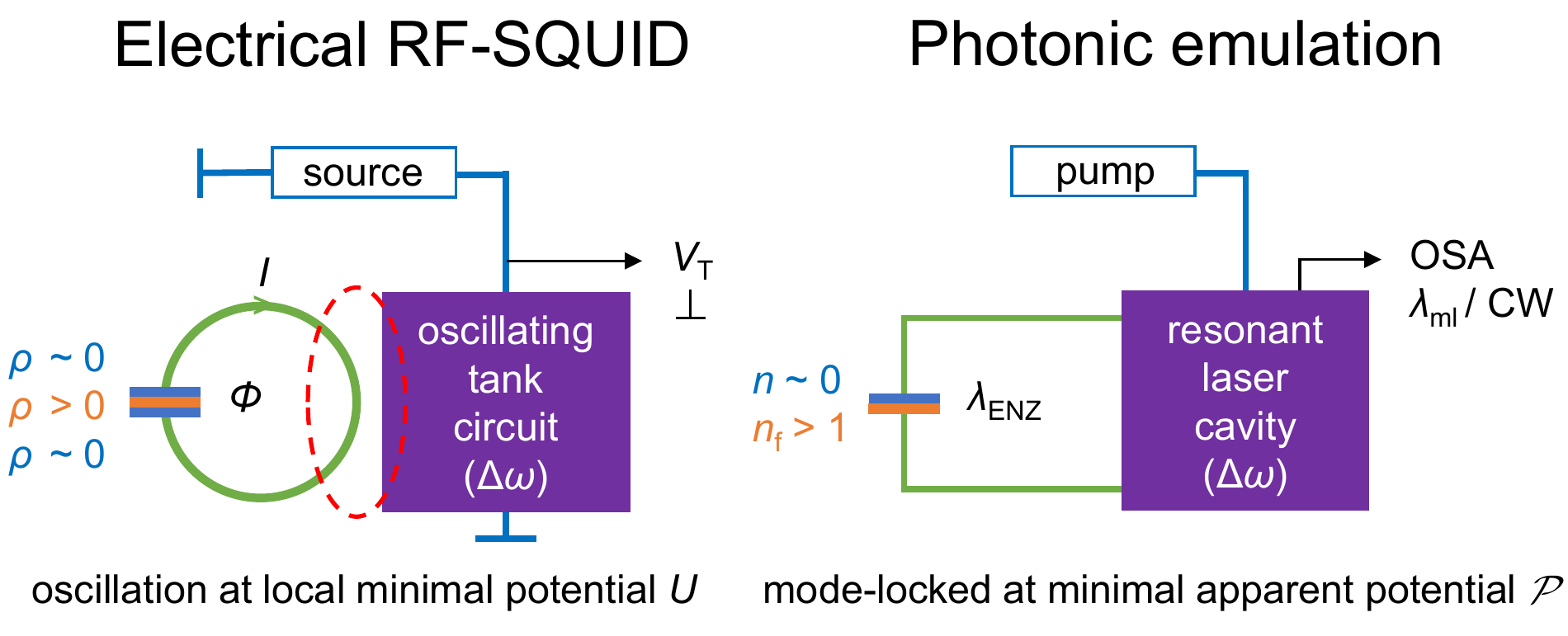} 
	\caption{The schemes of the electric RF-SQUID and its photonic counterpart. Note that in the dispersive mode of the operation of the electric SQUID, the resistance actually remains equal to zero in all components of the superconducting loop due to quantum tunneling.}
	\label{f1}
\end{figure}

A typical electric RF SQUID consists of an SC ring interrupted by a single JJ, coupled to an oscillatory TC \cite{Nisenoff1975}, as shown in Figure \ref{f1}. The JJ is an SC-dielectric-SC structure, and the TC can be constructed as a simple inductor-capacitor (LC) circuit, whose fundamental resonance frequency is $\omega_{0}=(LC)^{-1/2}$. When the RF-SQUID operates in the dispersive mode detecting a magnetic field, the induced current $I$ does not exceed the critical value $I_{\mathrm{c}}$. Then, the variation of the oscillation frequency in the tank circuit, $\Delta \omega $, can be detected by measuring the respective voltage (see Figure \ref{f1}), and a change in the magnetic field changes the inductance of JJ and leads to a change in the resonant frequency of the TC. With $I>I_{\mathrm{c}}$ (for a particular flux), the RF-SQUID operates in the resistive mode, and $I$ decays exponentially until the JJ recovers its superconducting state \cite{muck2001radio}. The TC must supply energy to the SQUID, which reduces the voltage across the TC, but does not necessarily change its oscillation frequency. In this mode, if only a small magnetic field is present, $I_{\mathrm{c}}$ would not be essentially exceeded.

\begin{figure*}[h]
	\centering
	\includegraphics[width=1\linewidth]{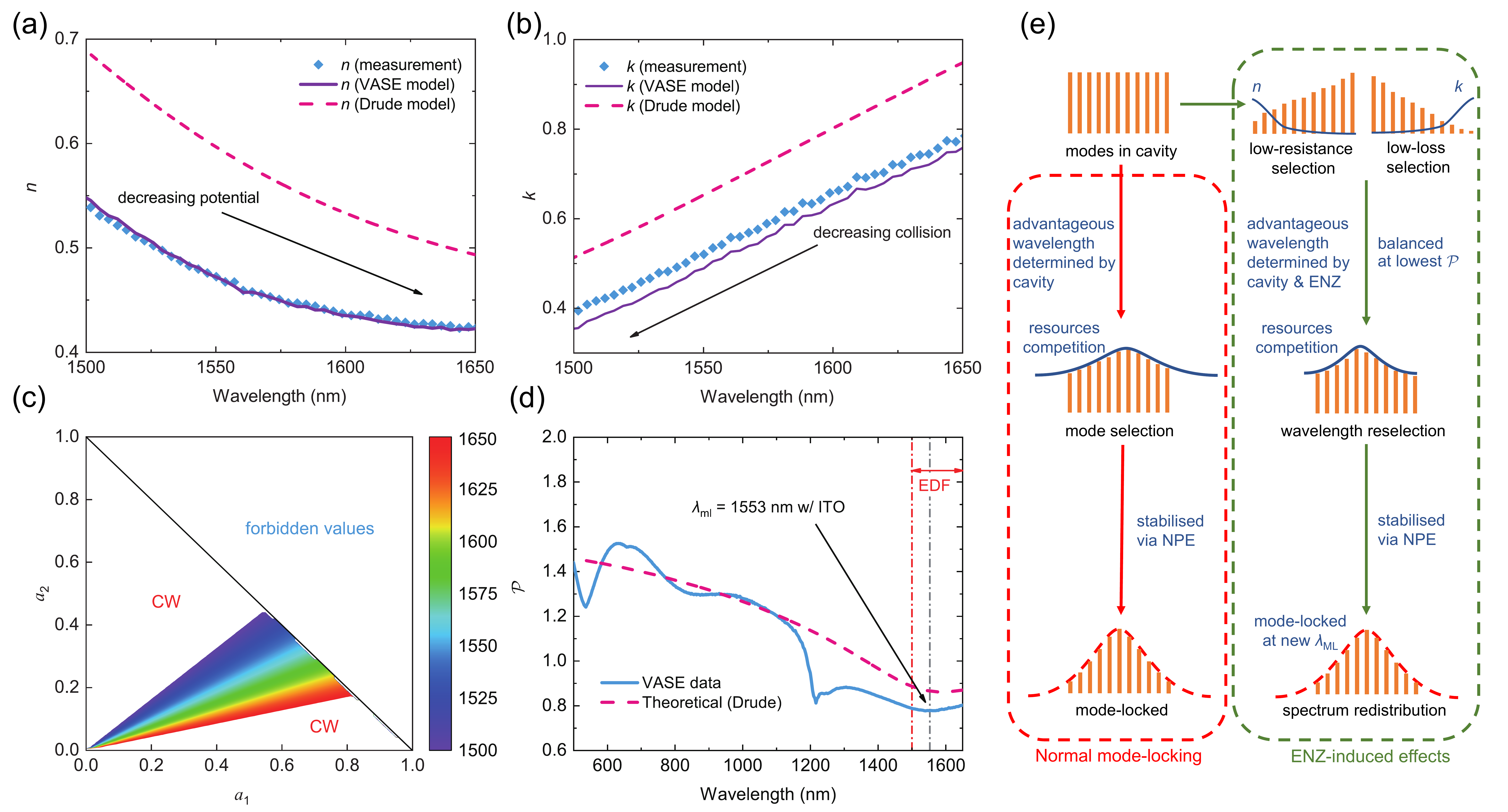} 
	\caption{The wavelength reselection and spectrum redistribution mechanism. a) and b), the real and imaginay parts of the refractive index, $n$ and $k$ (the extinction coefficient), vs. the wavelength. The experimental data and model produced by the VASE measurement, and their counterparts produced by the Drude model as per Eq. (\protect\ref{e2}), are presented, severally, by diamonds, solid and dashed lines. The arrows denote the low-potential bias or preference of the wavelength selection. c) The stable central wavelength-reselection area within the cavity's operation band with the lowest potential barrier. The color code corresponds to $\protect\lambda $ at the lowest $\mathcal{P}$ under the given weight coefficients $a_{1},a_{2},a_{3}$. d) The theoretical prediction of the effective height of the potential barrier, produced by the phenomenological model using the Drude permittivity dispersion, and its experimentally identified counterpart calculated from the VASE data. e) A schematic diagram of the mechanisms of the ENZ-induced wavelength reselection and spectrum redistribution.}
	\label{f2}
\end{figure*}

Thus, the RF-SQUID is built of two main components, the JJ serving as the sensor of the magnetic field, and TC that translates the detected field into a change of the oscillation frequency. It is worth mentioning that the RF-SQUID, with its intrinsic inductance, can operate even without a TC, but in this work we are considering the basic setup with a TC. In optics, the refractive index may be regarded as representing the ``resistance'' of the medium. Having near-zero permittivity and refractive index, ENZ materials realize ultra-low optical potential barriers, exerting low resistance onto the propagating optical wave. Therefore, this photonic setup resembles the SQUID scheme, with the ENZ element embedded in the laser cavity being a counterpart of the JJ, cf. in Figure \ref{f1}. It is mentioned in the \textit{Introduction} that the intra-cavity behavior of the ENZ element is different from that previously reported for extra-cavity setups. The basic reason for the difference is that, in the former case, light keeps traversing the ENZ in the course of roundtrips, until it is emitted out of the cavity. As a result, the effect of the wavelength/frequency preference accumulates. This wavelength-selecting feature of the optical cavity provides emulation of the TC function in the electric RF-SQUID, as sketched in Figure \ref{f1}. Resembling the RF-SQUID, this photonic counterpart of TC should also provide two operation modes. Namely, in the mode-locking regime the variation of $\lambda_{\mathrm{ENZ}}$ entails a change in the cavity’s resonant (mode-locking) frequency, as represented by $\lambda_{\mathrm{ML}}$. On the other hand, in the mismatched regime the variation of $\lambda_{\mathrm{ENZ}}$ keeps producing the CW output with an unchanged wavelength. A comparison table between the RF-SQUID and its optical counterpart is shown in the \textit{Supporting Information}.

\section{The Phenomenological Picture: Mode Reselection under the Action of the Optical-Potential-Barrier}

To realize the wavelength (frequency) shift within the cavity following the SC-dielectic analogue, \textit{i.e.}, in terms of the complex and rapidly varying index in the ENZ region, a mode-reselection-induced wavelength shift scheme should be utilized instead of the adiabatic \cite{Khurgin2020} and self-phase-modulation-induced ``time refraction'' \cite{Zhou2020}, \textit{i.e.}, a frequency shift between two segments of an optical medium separated by a boundary moving in time \cite{Plansinis2015}. Therefore, a weak-power nonlinear-polarization-evolution (NPE) cavity is considered, in which the added ENZ element should not greatly change the original mode-locking mechanism, rather acting as a mode reselector. Here, we propose a phenomenological picture based on the optical-potential-barrier concept to realize the mode reselection.

To demonstrate this more intuitively, we use the ellipsometry data from a 2 cm $\times $ 2 cm 300-nm thick ENZ indium tin oxide (ITO) nanolayer sample, which is fabricated on a pure silica substrate with the free-carrier concentration of $8.273\times 10^{20}$ cm$^{-3}$ and mobility 22.19 cm$^{2}$ (V$\cdot $s), exhibiting the ENZ point at $\lambda_{\mathrm{ENZ}}=1538$ nm where the real part of the permittivity vanishes (see \textit{Materials and methods}). The complex permittivity and the corresponding complex refractive index of the sample are measured by the variable-angle spectroscopic ellipsometer (VASE), with the measured and modeled index curves shown in Figures \ref{f2}(a) and \ref{f2}(b).

The complex permittivity curves of the ENZ ITO feature a trend similar to that in the basic Drude model. Therefore, to simplify the description of the proposed phenomenological model, we use the Drude model to illustrate the frequency dependence of the complex permittivity,

\begin{equation}
	\varepsilon ={\varepsilon_{\mathrm{r}}}+i{\varepsilon_{\mathrm{i}}}={\varepsilon_{\infty }}-\frac{{\omega_{\mathrm{p}}^{2}}}{{{\omega ^{2}}+{\gamma ^{2}}}}+i\frac{{\omega_{\mathrm{p}}^{2}\gamma }}{{\left( {{\omega^{2}}+{\gamma ^{2}}}\right) \omega }},
	\label{e1}
\end{equation}

\noindent where $\varepsilon_{\infty }$ is the high-frequency limit of the permittivity, the plasma frequency is $\omega_{\mathrm{p}}^{2}=Ne^{2}/(\varepsilon_{0}m^{\ast })$, and the damping rate is $\gamma =e/(\mu m^{\ast })$. Here, $N$ is the density of free carriers with effective mass $m^{\ast }$ and mobility $\mu$, while the frequency is related to the wavelength as usual, $\omega =2\pi c/\lambda $. In Eq. \ref{e1}, the complex permittivity can be converted to a complex refractive index, namely,

\begin{equation}
	{n_{\mathrm{C}}}=n+ik=\sqrt{\frac{{\sqrt{\varepsilon_{\mathrm{r}}^{2}+\varepsilon_{\mathrm{i}}^{2}}+{\varepsilon_{\mathrm{r}}}}}{2}}+i\sqrt{{n^{2}}-{\varepsilon_{\mathrm{r}}}}.
	\label{e2} 
\end{equation}

Due to the existence of $\varepsilon_{\mathrm{i}}\neq 0$, Eq. \ref{e2} always yields $n>0$. The dependence of $n$ and $k$ on the wavelength, corresponding to the measurements and models, is shown in Figures \ref{f2}(a) and \ref{f2}(b), respectively. Note that the numerical VASE model (solid lines), which takes into account all effects, agrees well with the measurements (diamonds), and demonstrates the same trend as the simple Drude model (the dashed lines).

The laser cavity with the population inversion, gain, and loss is a non-Hermitian optical system \cite{Sindelka2019}. In such a setting, the distribution of the refractive index may be treated as an effective optical potential \cite{ElGanainy2007}, following its quantum-mechanical counterpart, well known in the context of the \textit{parity-time symmetry} \cite{Bender1998}. The respective potential barrier affects the speed of light in the medium. In quantum mechanics, the wave function confined by potential barriers can form a trapped state. In optics, similarly, nanophotonic structures with periodic modulation of the refractive index can trap localized EM modes \cite{John1987, Yablonovitch1987}.

As seen from Figure \ref{f2}(a), the height of the optical potential barrier, determined by the real part $n$ of the refractive index, drops significantly at $\lambda >\lambda_{\mathrm{ENZ}}$. The distribution of $n$ is not even across the spectrum, therefore the EM waves with different values of $\lambda $ experience different ``resistance", with the longer-wavelength components having to pass lower potential barriers. This bias results in a wavelength preference, giving the longer-wavelength components an advantage (e.g., higher velocity and a shorter roundtrip time) in the competition for the optical gain. Additionally, varying $n$ rearranges the phase relation between the modes. Therefore, within the operation band of the cavity, the trend (shown in Figure \ref{f2}(a) by the arrow) is that the laser cavity with the inserted ENZ element mode-locks at $\lambda >\lambda_{\mathrm{ENZ}}$.

However, the presence of non-negligible intrinsic loss of the ENZ ITO sample makes it also necessary to consider the impact of the imaginary part of the refractive index (alias the extinction coefficient), $k$. In terms of effective complex potential barriers in non-Hermitian optics, $k$ determines results of the inelastic collision of the incident wave and barrier. The theoretical prediction and experimentally measured extinction coefficients are shown in Figure \ref{f2}(b).

On the contrary to the dependence $n(\lambda)$, $k$ rapidly increases as a function of the wavelength at $\lambda >\lambda_{\mathrm{ENZ}}$. For the longer wavelengths, the attenuation is so large that it outweighs the advantage provided by lower $n$. This trend is denoted by the arrow in Figure \ref{f2}(b).

The wavelength reselection and spectral redistribution also depend on properties of the cavity. To estimate the interplay of these different trends, inspired by the effective medium theory \cite{Felbacq1997} and its application on the weighted overall permittivity calculation of ENZ multilayer structures, we propose an empirical formula that comprises the total impact of all the factors:

\begin{equation}
	\mathcal{P}(\lambda )={a_{1}}\mathrm{Re}\left\{ {{n_{\mathrm{C}}(\lambda )}}\right\} +{a_{2}}\mathrm{Im}\left\{ {{n_{\mathrm{C}}(\lambda )}}\right\} +{a_{3}}{n_{\mathrm{f}}},
	\label{e3}
\end{equation}

\noindent where $\mathcal{P}$ is the effective potential height for wavelength $\lambda$, and, in our phenomenological model, $\lambda$ corresponding to the lowest $\mathcal{P}$ in the spectrum is the advantageous wavelength for the mode locking. In Eq. (\ref{e3}), $a_{1}$, $a_{2}$, and $a_{3}$ are weight coefficients that represent the influence of each factor. They are subject, by definition, to condition $a_{1}+a_{2}+a_{3}=1$. The first, second, and third terms in Eq. (\ref{e3}) are contributions to the overall potential height from $n$, $k$, and the cavity's structure. In particular, the presence of the cavity term $n_{\mathrm{f}}$ in Eq. (\ref{e3}) indicates that the combined potential barrier is most passable not exactly at $\lambda =\lambda_{\mathrm{ENZ}}$, but at the wavelength determined by the interplay of the ENZ element and the cavity. Therefore, the actual wavelength corresponding to the ML regime may be either $\lambda_{\mathrm{ML}}>\lambda_{\mathrm{ENZ}}$ or $\lambda_{\mathrm{ML}}<\lambda_{\mathrm{ENZ}}$. This value should be found from the minimum condition, $\mathrm{d}{\mathcal{P}}/\mathrm{d}\lambda =0$. The usual RF-SQUID also works in the regime realizing a minimum of the JJ potential barrier \cite{muck2001radio}.

Figure \ref{f2}(c) displays the stable wavelength-reselection region within the operation band of the cavity from Eq. (\ref{e3}), produced by sweeping $a_{1}$ and $a_{2}$. The forbidden values in Figure \ref{f2}(c) pertain to an area where $a_{1}+a_{2}+a_{3}>1$, which contradicts the definition of the weight coefficients. The CW-output regions are areas where $n$ and $k$ are ``over-dominant'', making $a_{3}<0$ at lowest $\mathcal{P}$, which implies that the hybrid mode locking is impossible. This typically happens when the ENZ range is located too far away from the laser-operation range. From Figure \ref{f2}(c), one can see that the proposed theory can predict a wide range of $\lambda_{\mathrm{ML}}$, which implicates the possibility for the proposed SQUID-like behavior. Further, Figure \ref{f2}(d) compares the effective height of the potential barrier, as obtained from the experimental data (for the sample with $\lambda_{\mathrm{ENZ}}=1538$ nm), and the theoretical prediction based on the Drude model.

In Figure \ref{f2}(d), the experimental values of the influence factors from Eq. (\ref{e3}) are $a_{1}=0.51$ and $a_{2}=0.19$. The theoretically predicted optimal regime coincides with the experimental data within the cavity's operation band, the respective value of $a_{2}$ being $12.5\%$ larger than the experimental one, which qualitatively verifies the phenomenological consideration. Once the dominant wavelength is selected by the cavity, sidebands form as they would in the cavity without the ENZ element, thus finalizing the spectrum redistribution. The buildup of the spectrum is illustrated in Figure \ref{f2}(e).

\begin{figure*}[th]
	\centering
	\includegraphics[width=0.7\linewidth]{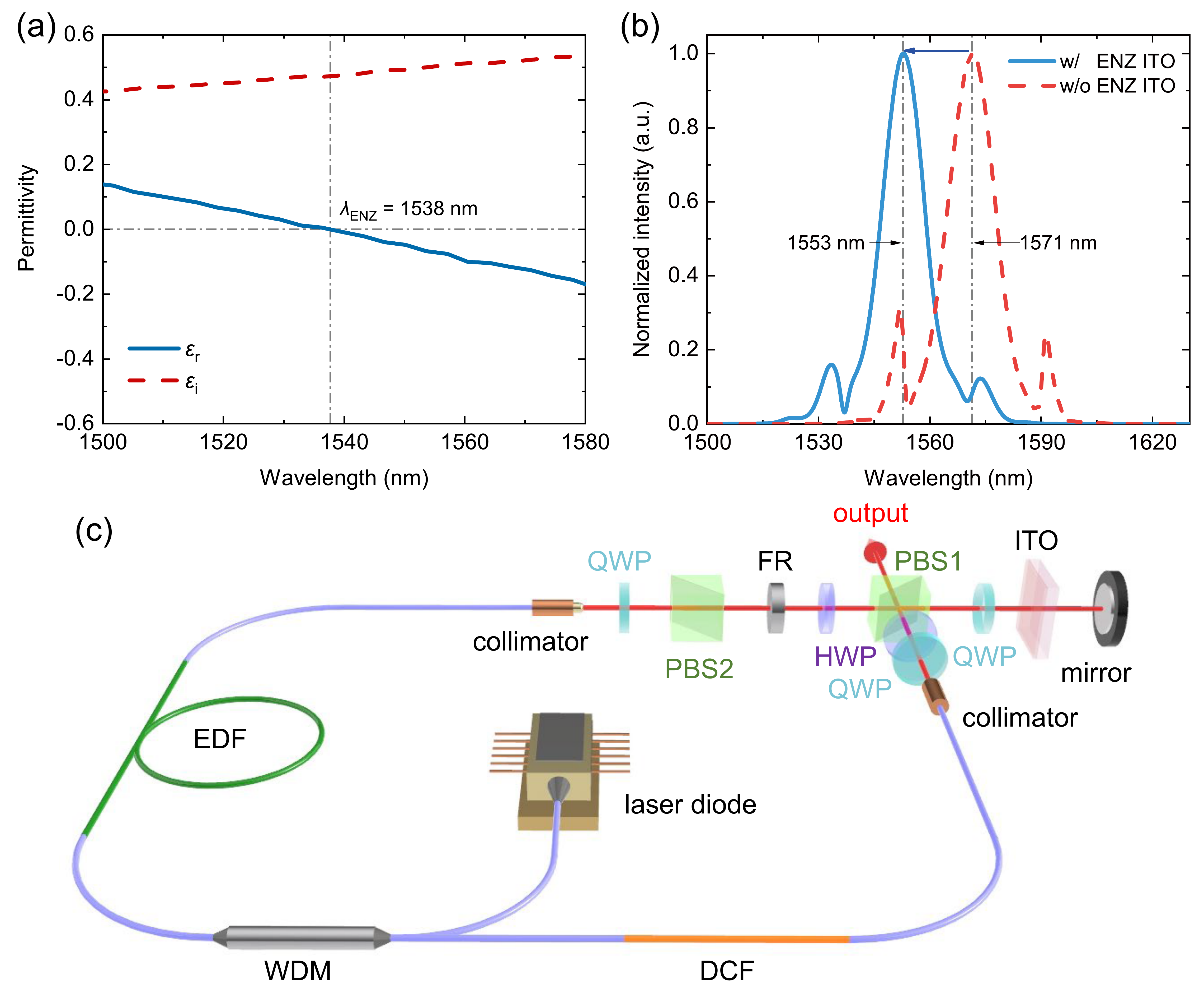} 
	\caption{The realization and observation of the SQUID-like behavior in the NPE cavity. a) The real and imaginary (solid and dashed lines) parts of the ENZ element of the cavity from the VASE measurement. b) The output spectra of the cavity with/without (solid/dashed lines) the inserted ENZ ITO sample. c) The experimental scheme. DCF: dispersion compensation fiber; EDF: erbium-doped fiber, FR: Faraday rotator; PBS: polarization beam splitter; WDM: wavelength division multiplexer; WP: waveplate (HWP, half WP; QWP, quarter WP).}
	\label{f3}
\end{figure*}

When the ENZ condition is no longer satisfied within the considered band, the benefit of low $n$ vanishes, which results in phase accumulation and creates mismatch in the polarization evolution. Therefore, the mode locking is lost if a pure glass substrate, or an ITO sample whose ENZ out of the laser operating range, is inserted in the cavity. Also, in the latter case, the extinction factor $k$ is relatively large, as seen in Figure \ref{f2}(b), hence the strong absorption further destroys the ML regime at all wavelengths. Thus, the proposed phenomenological picture strongly suggests that the role of ENZ ITO is that its highly dispersive real and imaginary parts of the permittivity act as a mode re-selector in the cavity while preserving the NPE condition.

\section{The Wavelength Reselection and Spectrum Redistribution}

Here, we realize the proposed photonic SQUID-like setup in the experiment, which makes it necessary to vary the optical wavelength/frequency in the resonant laser cavity. If the ENZ sample is absent, the cavity operates in its ML state at wavelength $\left( \lambda_{\mathrm{ML}}\right)_{0}$ (corresponding to the original resonant frequency, $\omega_{0}$). If the sample's ENZ wavelength is present in the laser operation range, a shift in $\lambda_{\mathrm{ML}}$ should be observed. Similar to the action of the real SQUID, which detects the external magnetic field and converts it into the electric signal, the photonic emulator ``detects'' the presence of the ENZ and translates it into a shift of the laser emission, as shown in Figure \ref{f1}. In this Section, we use the same sample mentioned in the phenomenological model, whose ENZ characteristics are shown in Figure \ref{f3}(a).

To realize the ENZ internal feedback, we designed the experimental setup displayed in Figure \ref{f3}. The laser configuration consists of different fiber pieces, fiber pigtailed optical components, and free-space optics for the ML of the NPE. To incorporate the sample in the free space of the laser cavity, we use the $\sigma $-shaped design of the fiber laser, as shown in Figure \ref{f3}(c). A branch of the contour is realized at PBS1, where one of the beams is directed to the ENZ sample perpendicularly, and the transmitted light is reflected by the highly reflective silver mirror, passing through the ENZ nanolayer reversely, getting back into the cavity. This design allows the removal of ENZ ITO sample at any time, and the cavity reverting back to the conventional mode-locked operation without any change and recalibration of the optical path.

Without the ENZ ITO sample, the loop cavity stably yields a $57$ MHz train of $15$ nm wide (full width at half maximum) $250$-fs ultrashort pulses centered at $1571$ nm (see \textit{Supporting Information}). When the ENZ ITO sample with $\lambda_{\mathrm{ENZ}}=1538$ nm is inserted, the center wavelength shifts to $1553$ nm, as shown in Figure \ref{f3}(b). The shift is $\approx 120\%$ of the spectral width of the pulse, while the pulse's spectral and temporal shapes remain almost unchanged. Thus, the stably reproducible spectrum is redistributed around the new central wavelength (see \textit{Supporting Information}). This effect persists when different wavelengths of the light source $(\lambda_{\mathrm{ML}})_{0}$ are used, see \textit{Supporting Information}.

\begin{figure*}[th]
	\centering
	\includegraphics[width=0.7\linewidth]{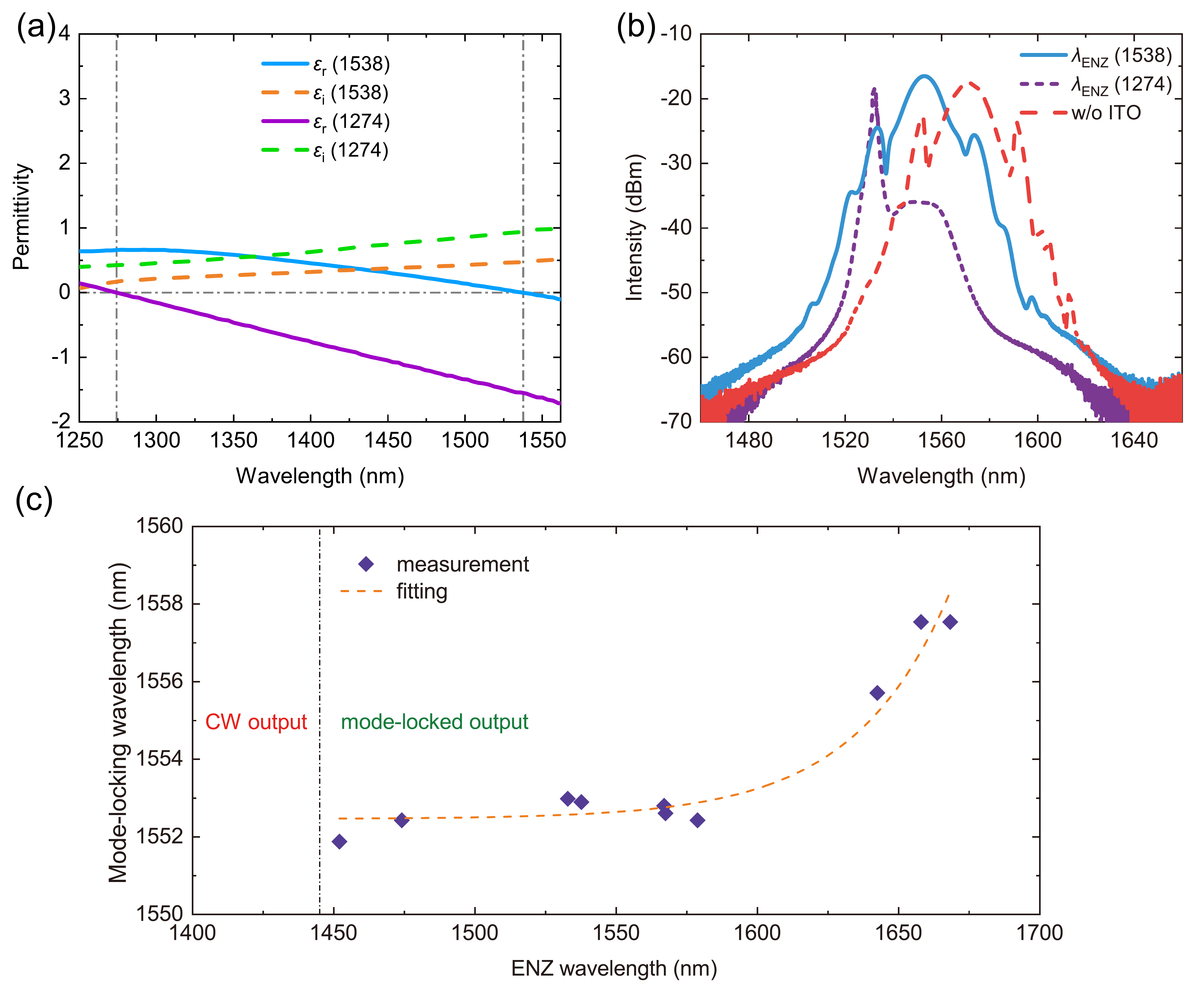} 
	\caption{The comparison between the results for different ENZ ITO samples. a) The real and imaginary parts of the permittivity for $\protect\lambda_{\mathrm{ENZ}}=1538$ and $1274$ nm. b) The output spectra shown on the logarithmic scale for these two samples and, in addition, for the setup without the ENZ ITO sample (the long-dashed line). c) The relation between values of $\protect\lambda_{\mathrm{ENZ}}$ and the corresponding $\protect\lambda_{\mathrm{ML}}$, as obtained from measurements performed with twelve different ITO samples, having $\protect\lambda_{\mathrm{ENZ}}=$ 1274.15, 1434.00, 1451.99, 1474.05, 1532.84, 1537.71, 1566.95, 1567.49, 1578.83, 1642.48, 1657.92, and 1668.29 nm. The first two samples (with $\protect\lambda_{\mathrm{ENZ}}=1274.15$ and $1434.00$ nm) are out of the laser's operation range, producing CW outputs, which have no corresponding $\protect\lambda_{\mathrm{ML}}$.}
	\label{f4}
\end{figure*}

Note that, by adjusting the waveplates (WPs), the central wavelength of the pulse can also be shifted. However, in this experiment the only difference between observations is produced by the insertion/removal of the ENZ ITO sample, \textit{without recalibration and readjustment of the WPs}. Therefore, the spectrum shift is induced by the ITO alone, which agrees well with the predicted behavior by the phenomenological model.

\section{Further Assessment of the Intra-Cavity ENZ Effect}

Although thin-film ENZ elements are used in many optical setups \cite{Guo2017, Zhou2020}, the role of ENZ in this work is altogether different. As many factors could potentially influence and/or contribute to the observed phenomenon, we analyze them as follows, noting the ability of ITO to switch its characteristics by means of saturable absorption \cite{Jiang2018, Xiao2020, Zhang2020}. In our experiments, the nonlinear saturable absorption is also observed (see \textit{Supporting Information}), which may contribute constructively to the hybrid mode locking. However, if the optical-switching effect of the ITO film indeed plays the main role, the laser is expected to achieve the ML state regardless of $\lambda_{\mathrm{ENZ}}$. The nonlinear Kerr effect, on the other hand, is trivial in this context, due to the weak intensity in the fiber laser cavity. Additionally, using the two-temperature model \cite{Alam2016, Alam2018}, the calculation of the intensity-induced ENZ wavelength shift shows that it is smaller than $0.1$ nm, which may be neglected.

To corroborate the validity of the operation regime outlined above and rule out essential effects produced by factors unrelated to the ENZ effect, in this Section we switch to the sample with $\lambda_{\mathrm{ENZ}}=1274$ nm, placed far outside the operation band of the cavity. The comparison of the complex permittivity of the two samples measured by VASE is presented in Figure \ref{f4}(a). It is seen that the sample with $\lambda_{\mathrm{ENZ}}=1274$ nm features negative real permittivity and a relatively high intrinsic loss (the imaginary part) within the cavity's operation band, while the sample with $\lambda_{\mathrm{ENZ}}=1538$ nm exhibits ENZ and lower loss in the considered band.

\begin{table*}[!ht]
	\caption{Comparison of intracavity-ENZ-based laser systems }
	\tiny
	\resizebox{\linewidth}{!}{
		\begin{tabular}{lllllllll}
			\toprule
			Ref. & Operation & \makecell[l]{ENZ \\component} & \makecell[l]{$\lambda_{\mathrm{ENZ}}$ \\(nm)} & \makecell[l]{$\lambda_{\mathrm{output}}$ \\(nm)} & \makecell[l]{Pulse \\width} & \makecell[l]{Repetition \\rate} & $P_{\mathrm{avg}}$ /  $P_{\mathrm{pump}}$ & Features \\ \midrule
			\multirow{2}*{\makecell{This \\work}} & \makecell[l]{mode-locked \\by NPE} & none & none & 1571 & \multirow{2}*{250 fs} & \multirow{2}*{57 MHz} & 10.47 mW / 130 mW & \makecell[l]{$\lambda_{\mathrm{ML}}$ can be tuned by WPs. \\CW output at NPE mismatch.} \\  \cline{2-5} \cline{8-9}
			& hybrid mode-locked & ITO thin film & 1451.99--1668.29 & 1551.88--1557.54 & & & 11.82 mW / 240 mW & \makecell[l]{$\lambda_{\mathrm{ML}}$ can be tuned by $\lambda_{\mathrm{ENZ}}$. \\CW output at $\lambda_{\mathrm{ENZ}}$ mismatch.} \\ \midrule
			\cite{Guo2017} & \makecell[l]{mode-locked \\by transient bleach} & ITO nanocrystal & $\approx$1300--1600 & $\approx$ 1560 & $\approx$ 593 fs & 16.62 MHz & 0.26 mW / 12 mW & \makecell[l]{Sample exhibits large modulation \\depth and sub-ps response time}\\ \midrule
			\cite{Jiang2018} & \multirow{3}*{\makecell[l]{Q-switched \\by \\saturable absorption}} & ITO thin film & $\approx$2000 & 1862.5 & 526--882 ns & 241--113 kHz & 97.2 mW / 700 mW & \multirow{3}*{\makecell[l]{Demonstration of \\ENZ-based Q-switched lasers \\outside C-band}}\\ \cline{1-1} \cline{3-8}
			\cite{Xiao2020} & &  ITO nanocolumns array & 1293.5 & 1064.63 & 579.6--1060 ns & 126.1--67.4 kHz & 152 mW / 1.85 W & \\ \cline{1-1} \cline{3-8}
			\cite{Zhang2020} & & ITO with nanostructures & N. A. & 2062.8 & 2.42 $\mu$s & 20.53 kHz & 312 mW / 2.56 W & \\ \bottomrule
	\end{tabular}}
	\label{t1}
\end{table*}

Output spectra of the cavity with the two different ITO samples are displayed in Figure \ref{f4}(b). One can see that the output of the sample with $\lambda_{\mathrm{ENZ}}=1274$ nm has the CW form, without any signature of the ML regime. In this case, no phase-locked spectrum can be observed, no matter how much pump energy is applied. On the contrary to the previous situation, only high-intensity CW appears at wavelength $1532$ nm, and the laser fails to reach a new steady state for spectral reconstruction and phase-locking. This can be explained by the fact that the ITO's switching effect is not the cause of the appearance of the spectrum shown in Figure \ref{f3}(b), and, within the considered operation band, the sample with $\lambda_{\mathrm{ENZ}}=1274$ nm exhibits no ENZ effect (\textit{i.e.}, the refractive index in this range is no longer close to zero). With no adjustment of the optical path, the output may only take the CW form, and even the saturable absorption alone cannot realize the WP-adjustment-free hybrid mode-locking. This behavior further demonstrates that the wavelength reselection and spectrum redistribution are caused by the unique ENZ effects.

Additionally, to evaluate the impact of $\lambda_{\mathrm{ENZ}}$ on the actual value of $\lambda_{\mathrm{ML}}$, the experiment was reproduced with a set of twelve pieces of ITO samples with different values of $\lambda_{\mathrm{ENZ}}$ inserted in the cavity (see \textit{Supporting Information}). The preparation of the sample is technically complex and time-consuming, because tuning $\lambda_{\mathrm{ENZ}}$ from $1274$ nm across the C-band to $1668$ nm is a challenging procedure. The results are displayed in Figure \ref{f4}(c), which shows a trend for longer $\lambda_{\mathrm{ENZ}}$ to have a greater effect on the wavelength reselection, especially for $\lambda_{\mathrm{ENZ}}>\left(\lambda_{\mathrm{ML}}\right)_{0}=1571$ nm. It is worthy to note that the established value of $\lambda_{\mathrm{ML}}$ is not necessarily located between $\lambda_{\mathrm{ENZ}}$ and original $\lambda_{\mathrm{ML}}$, which is explained by the ML mapping in Figure \ref{f4}(c). In line with these considerations, for wavelengths $\lambda_{\mathrm{ENZ}}$ which are too short (\textit{viz.}, $1274$ and $1434$ nm), lying outside of the operation range of the cavity, a CW output is obtained.

The relation between $\lambda_{\mathrm{ENZ}}$ and $\lambda_{\mathrm{ML}}$ in Figure \ref{f4}(c) indicates a ``detection capability'' of the ENZ-based setup, based on the wavelength reselection and spectrum redistribution. On the other hand, similar to the resistive mode in the RF-SQUID, with $I>I_{\mathrm{c}}$ and decaying $I$, here the CW output persists at $\omega_{\mathrm{ENZ}}>\omega_{\mathrm{gain,min}}$, when the laser cavity cannot operate in the ML regime. The relation of $\lambda_{\mathrm{ENZ}}$ to $\lambda_{\mathrm{ML}}$, shown in Figure \ref{f4}(c), unveils a novel method to directly identify an initially unknown $\lambda_{\mathrm{ENZ}}$ by putting it into the cavity and comparing the results with those obtained for a known $\lambda_{\mathrm{ENZ}}$, without referring to expensive and non-ubiquitous ellipsometry measurements. This method can be utilized for designing new instruments for optical measurements.

Conclusively, considered from the perspective of the system, rather than the ENZ material itself, there are fundamental differences between the observed SQUID-like behavior and previous extra-cavity frequency-shift experiments (see, \textit{e.g.}, Refs. \cite{Khurgin2020, Zhou2020}), namely, i) the number of wavelength components inside the cavity is more diverse than in extra-cavity settings, with the former arrangement exhibiting the mode selection, competition for gain resources, and traveling through ENZ TCO as the additional selection condition in multiple roundtrips; ii) the temporal overlap of pulses in a pulse train in the laser cavity is not guaranteed to initiate a pump-probe relation in the time-refraction frequency shift; iii) due to the weak power level, the size of the calculated Kerr-induced refractive-index change is only $1.8\%$ at the ENZ wavelength (from $0.63$ to $0.6413$), and the corresponding permittivity changes from $0$ to $0.0226$, which is insignificant. Thus, the single crucial factor is, indeed, is the rapidly changing complex permittivity, which shapes the unique dispersion and absorption curves of ITO in the ENZ region.

If the SQUID-like operation in the NPE laser cavity is considered as a whole, the following inferences can be made. For an NPE laser without ENZ, the polarization components form an artificial saturable absorber. The optical modes launched by the CW source propagate, while their polarizations rotate and evolve. Those with the highest transmission for the given polarization will resonate and lase. In such a system, the aforementioned exclusive dispersion and absorption curves of ITO in the ENZ region act as the mode-reselector, maintained by the original NPE mode-locking regime. These inferences confirm the SQUID-like behavior predicted by the underlying assumption. 

Additionally, it is worth mentioning that the response of a SQUID is periodic in the magnetic field. As the field is increased, the TC frequency varies periodically between a maximum and minimum value. Our proposed ENZ optical system is limited by the operation range of the laser (which in turn, is limited by the gain bandwidth of the erbium-doped fiber), therefore, just like SQUID, it has upper and lower limits of the frequency. The periodicity is not observed in our experiments, due to the fact that the ranges of $\lambda_{\mathrm{ENZ}}$ and the laser operation range have similar limited scales. Finally, to give a clear picture of the development of the intracavity ENZ studies, we further summarize a comparison of intracavity-ENZ-based laser systems in Table \ref{t1}. As can be concluded from the table, our SQUID-like ENZ laser setup provides wavelength tunability by switching ENZ samples, as well as a shorter pulse width and a higher repetition rate; while the Q-switched systems generally exhibit higher efficiencies than the mode-locked ones.

\section{Conclusions}

In this work, we have proposed and phenomenologically substantiated the novel RF-SQUID-like operation regime in the fiber laser with the intra-cavity ENZ element. The predicted behavior has been realized in the experimental setup. With $\lambda_{\mathrm{ENZ}}$ taken in the range of the cavity's operation band, different values of $\lambda_{\mathrm{ENZ}}$ yield differently-shifted mode-locked pulses, created by the wavelength reselection and spectrum redistribution in the setup. On the other hand, only the CW output occurs in the absence of the ENZ element under the action of the added-phase-induced polarization-evolution mismatch. The balance between the (small) real and imaginary parts of the effective complex optical potential barrier, and contributions to it from intrinsic characteristics of the laser cavity, determine the observed wavelength reselection and spectrum redistribution scenarios. This approach offers a new method for identifying $\lambda_{\mathrm{ENZ}}$, without using high-precision instruments. The results of this work offer a deeper insight into ENZ photonics, as well as additional tools for the investigation of superconductivity by means of photonic analogues.

\section*{Supporting Information}

\small{Supporting Information is available as a separate file.}

\section*{Acknowledgements}

\small{
This work is supported by Guangdong Basic and Applied Basic Research Foundation (Grant No. 2021A1515012176, 2021A1515011450), Youth Science and Technology Innovation Talent of Guangdong Province (Grant No. 2019TQ05X227), Shenzhen Fundamental Research Program \\(Grant No. GXWD20201231165807007-20200827130534001), Overseas Research Cooperation Fund of Tsinghua Shenzhen International Graduate School (Grant No. HW2020006), and Swiss National Science Foundation (Grant No. 200021\_188605). The work of B.A.M. is also supported, in a part, by the Israel Science Foundation through Grant No. 1286/17. 
}

\section*{Conflict of Interest}

\small{The authors declare no conflict of interest.}

\section*{Data Availability Statement}
\small{The data that support the findings of this study are available from the \textit{Supporting Information}.}

\section*{Keywords}
\small{Epsilon-near-zero, indium tin oxide, fiber-laser, mode-locking, nonlinear polarization evolution}

\bibliography{References}
\bibliographystyle{lpor}

\end{document}


\captionsetup[figure]{labelfont={bf},name={Figure},labelsep=none}
\flushbottom
\maketitle
\linespread{1.5}
\renewcommand\thesection{S\arabic{section}}
\renewcommand\theequation{S\arabic{equation}}
\renewcommand\thetable{S\arabic{table}}
\renewcommand\thefigure{S\arabic{figure}}
%
%
\clearpage

\section{RF-SQUID and fiber laser with SQUID-like behavior}
\label{SI00}

\begin{table*}[!ht]
	\caption{Comparison between the features of RF-SQUID and SQUID-like fiber laser}
	\tiny
	\resizebox{\linewidth}{!}{
		\begin{tabular}{llllllll}
			\toprule
			Device & \makecell[l]{Key \\component} & \makecell[l]{Oscillating \\component} & \makecell[l]{Physical quantity \\sensed} & \makecell[l]{Readout parameter} & \makecell[l]{Operation modes} & \makecell[l]{Mode behavior} & \makecell[l]{Alternative operation \\without oscillating components} \\ \midrule
			\multirow{2}*{RF-SQUID} & \multirow{2}*{\makecell[l]{Josephson junction \\of dielectric-SC}} & \multirow{2}*{tank circuit (TC)} & \multirow{2}*{\makecell[l]{magnetic flux \\($\Phi$)}} & \multirow{2}*{\makecell[l]{oscillation frequency in TC\\(dispersive mode)}} & dispersive mode & \makecell[l]{translate $\Phi$ to frequency \\variation ($\Delta \omega$) in TC} & \multirow{2}*{\makecell[l]{detect $\Phi$ by measuring \\the inductance of SQUID}} \\  \cline{6-7}
			& & & & & resistive mode & exponential decay of current & \\ \midrule
			\multirow{2}*{\makecell[l]{SQUID-like \\fiber laser \\setup}} & \multirow{2}*{\makecell[l]{ENZ-non-ENZ \\interface}} & \multirow{2}*{NPE laser cavity} & \multirow{2}*{\makecell[l]{ENZ wavelength \\($\lambda_{\mathrm{ENZ}}$)}} & \multirow{2}*{\makecell[l]{mode-locking wavelength \\($\lambda_{\mathrm{ML}}$) in laser cavity \\(mode-locking regime)}} & mode-locking regime & \makecell[l]{translate $\lambda_{\mathrm{ENZ}}$ to \\wavelength variation ($\lambda_{\mathrm{ML}}$) \\in laser cavity} & \multirow{2}*{\makecell[l]{detect $\lambda_{\mathrm{ENZ}}$ by measuring \\the permittivity using ellipsometer.}} \\ \cline{6-7}
			& & & & & CW regime & CW output & \\\bottomrule
	\end{tabular}}
	\label{t0}
\end{table*}

\section{Experimental methods}
\label{SI0}

\subsection{Sample fabrication} 

The ENZ ITO samples were fabricated by 100 W DC magnetron sputtering technology using a 99.99\% purity 10 wt\% In:Sn$_{2}$O$_{3}$ target. The samples were then annealed in vacuum for 2 hours at 350$^{\circ }$C.

\subsection{Experimental equipment and instruments} 

The complex permittivity was measured by an M-2000UI variable-angle spectroscopic ellipsometer (VASE) by J.A. Wollam Co. A YOKOGAWA AQ6370D optical spectrum analyzer (OSA) was used to monitor the spectral profiles. The electrical properties of the samples were measured by the Ecopia HMS-3000 Hall-effect measurement system. The data presented in the \textit{Supporting Information} were collected by means of a 3 GHz bandwidth InGaAs photodetector, a Keysight PXA Signal Analyser, a Femtochrome FR-103XL/IR/FA autocorrelator, and an InfiniiVision DSOX6004A oscilloscope from Keysight.

\section{Drude model parameters for ENZ ITO}
\label{SI1}

As said in the main text, the complex permittivity of the ENZ ITO with $\lambda _{\mathrm{ENZ}}=1538$ nm can be described by the Drude model \cite{Drude1900}:

\begin{equation}
	\varepsilon = {\varepsilon _{\mathrm{r}}} + i{\varepsilon _{\mathrm{i}}} = {\varepsilon _\infty } - \frac{{\omega _{\mathrm{p}}^2}}{{{\omega ^2} + {\gamma ^2}}} + i\frac{{\omega _{\mathrm{p}}^2\gamma }}{{\left( {{\omega ^2}+
				{\gamma ^2}} \right)\omega }},  
	\label{S1}
\end{equation}

Here, we calculate the Drude parameters, based on the ellipsometry experimental data, as follows (assuming $m^{\ast }=0.38m_{0}$) \cite{Naik2013, Niu2018}:

\vspace{8pt}

$\varepsilon_\infty = 4.4755$;

\vspace{8pt}

$\omega_{\mathrm{p}} = 2.6287\times10^{15}$ rad/s;

\vspace{8pt}

$\gamma = 2.0830\times10^{14}$ rad/s.

\section{The scheme of the mode-locking operation}
\label{SI2}

To achieve stable ML (mode-locking) of the nonlinear-polarization-evolution-based fiber laser, the following steps are required. Firstly, the combination of a polarization beam splitting (PBS), Faraday rotator (FR) and a half-wave plate (HWP) act as a Faraday isolator, and the HWP needs to be rotated by a certain angle to ensure the unidirectional operation. Secondly, randomly rotating the rest wave plates leads to an ML spectrum containing a continuous wave (CW) or multiple pulses. Thirdly, the optimization of pump power and the wave plate rotations are conducted to obtain clean and stable ML pulses. Once the ML behavior is established, directly ramping up the pump power can achieve self-starting and ML can be stably reproduced in the course of several days of the work.

\vspace{8pt}

Parameters of the cavity fibers are listed as follows:

\vspace{8pt}

DCF (0.2 m): $-160.72$ ps/nm/km, YOFC, DM1010-D

\vspace{8pt}

EDF (1 m): 15.9927 ps/nm/km, LIEKKI, Er80-8/125, Absorption: 80 dB/m at 1530 nm

\vspace{8pt}

SMF (2.2 m): 18 ps/nm/km

\section{Optical characteristics of ultrashort pulses with or without ENZ ITO nanolayer}
\label{SI3}

\begin{figure}[h]
	\centering
	\includegraphics[width=1\linewidth]{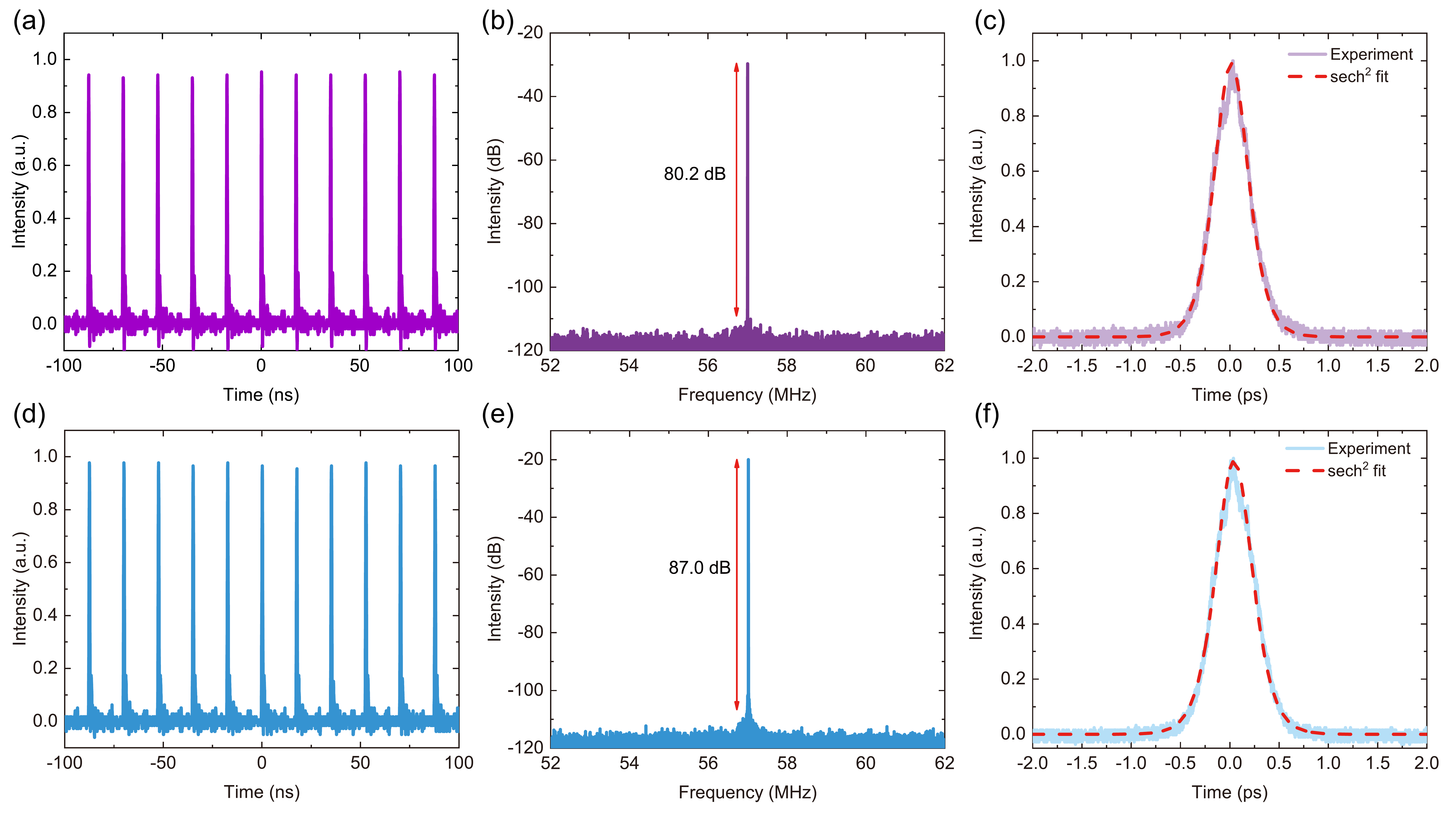} 
	\caption{. Temporal characteristics of the laser output with and without ENZ ITO nanolayer. Temporal profiles with ITO are in the top row, and those without ITO are in the bottom.}
	\label{fs1}
\end{figure}

Several optical parameters have been characterized in the measurement of the generated pulses. The pulse train with a temporal interval of 17.5 ns corresponds to a radio frequency (RF) spectrum centered at 57 MHz. The signal-to-noise ratio (SNR) of better than 80 dB indicates the superior pulse-to-pulse stability. The autocorrelation trace is fitted using a sech$^{2}$ function. All these characteristics are shown in Fig. \ref{fs1}. The decrease in signal-to-noise ratio (SNR) is due to the increment of relative intensity noise, shown in Section \ref{SI5}.

\section{The ENZ wavelength versus the ML wavelength}
\label{SI4}

To demonstrate the relation between the ML wavelength ($\lambda _{\mathrm{ML}}$) and the ENZ wavelength ($\lambda _{\mathrm{ENZ}}$), we have fabricated 12 samples with different $\lambda _{\mathrm{ENZ}}$. The data displayed in Fig. 2c of the main text are collected in the following Table \ref{t1}. 

\begin{table}[!h]
	\caption{The relation between the ML wavelength ($\lambda _{\mathrm{ML}}$) and the ENZ wavelength ($\lambda _{\mathrm{ENZ}}$).}
	\resizebox{\linewidth}{!}{\centering
		\begin{tabular}{ccccccccccccc}
			\midrule No. & 1 & 2 & 3 & 4 & 5 & 6 & 7 & 8 & 9 & 10 & 11 & 12 \\ \midrule
			$\lambda_{\mathrm{ENZ}}$ (nm) & 1274.15 & 1434.00 & 1451.99 & 1474.05 & 1532.84 & 1537.71 & 1566.95 & 1567.49 & 1578.83 & 1642.48 & 1657.92 & 1668.29 \\
			$\lambda_{\mathrm{ML}}$ (nm) & CW & CW & 1551.88 & 1552.43 & 1552.98 & 1552.90 & 1552.80 & 1552.61 & 1552.43 & 1555.71 & 1557.54 & 1557.54 \\
			\midrule
		\end{tabular}
	}
	\label{t1}
\end{table}

\section{Relative intensity noise with or without ENZ ITO nanolayer}
\label{SI5}

\begin{figure}[h]
	\centering
	\includegraphics[width=0.38\linewidth]{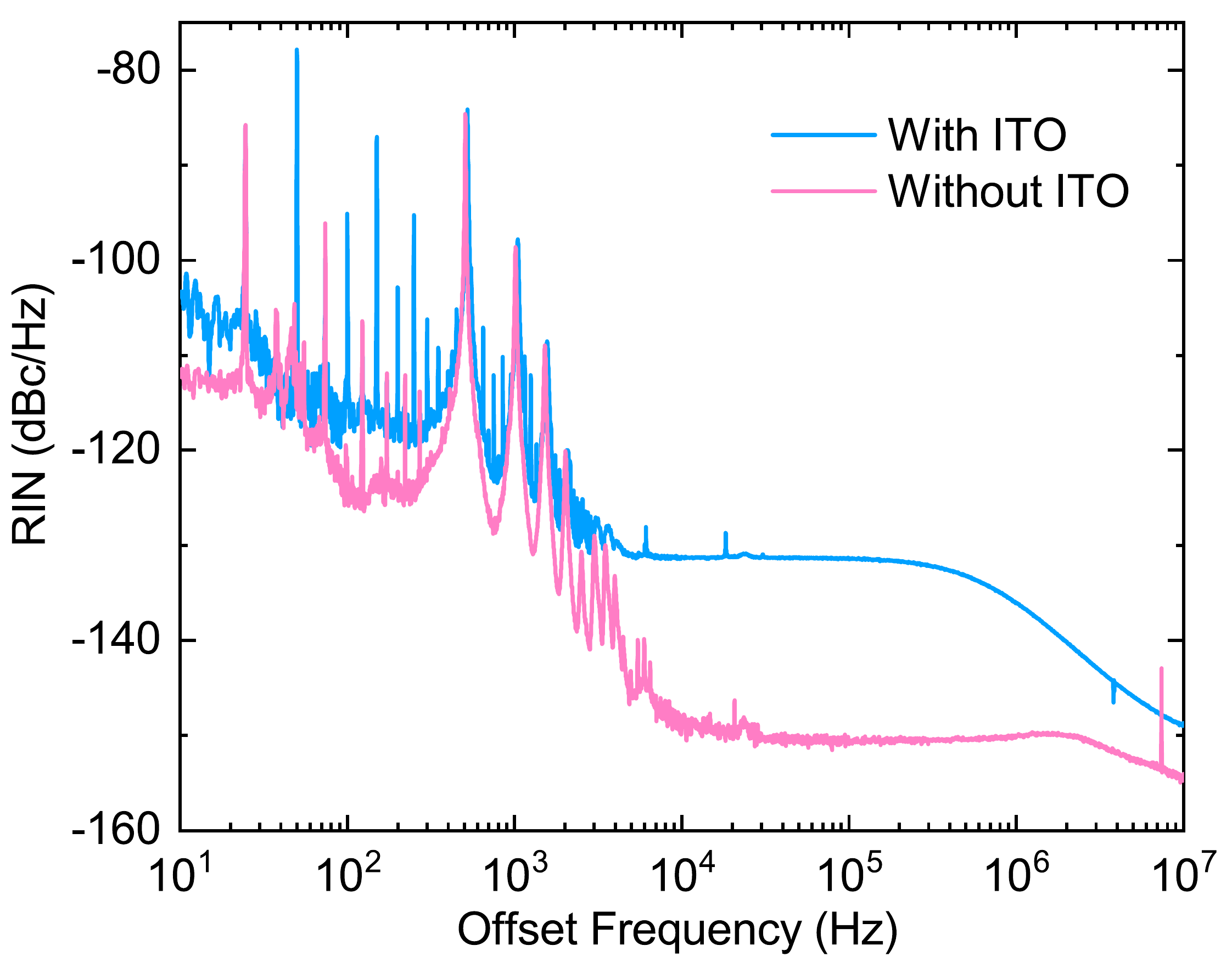} 
	\caption{. Relative intensity noise with and without ENZ ITO.}
	\label{fs2}
\end{figure}

The relative intensity noise (RIN) with and without ENZ ITO is measured using Rohde \& Schwarz FSWP8 phase noise analyzer, and the results are shown in Fig. \ref{fs2}. The existence of ITO increases the RIN of the laser output.

\section{Nonlinear saturable absorption characteristics of ENZ ITO}
\label{SI6}

\begin{figure}[!h]
	\centering
	\includegraphics[width=0.40\linewidth]{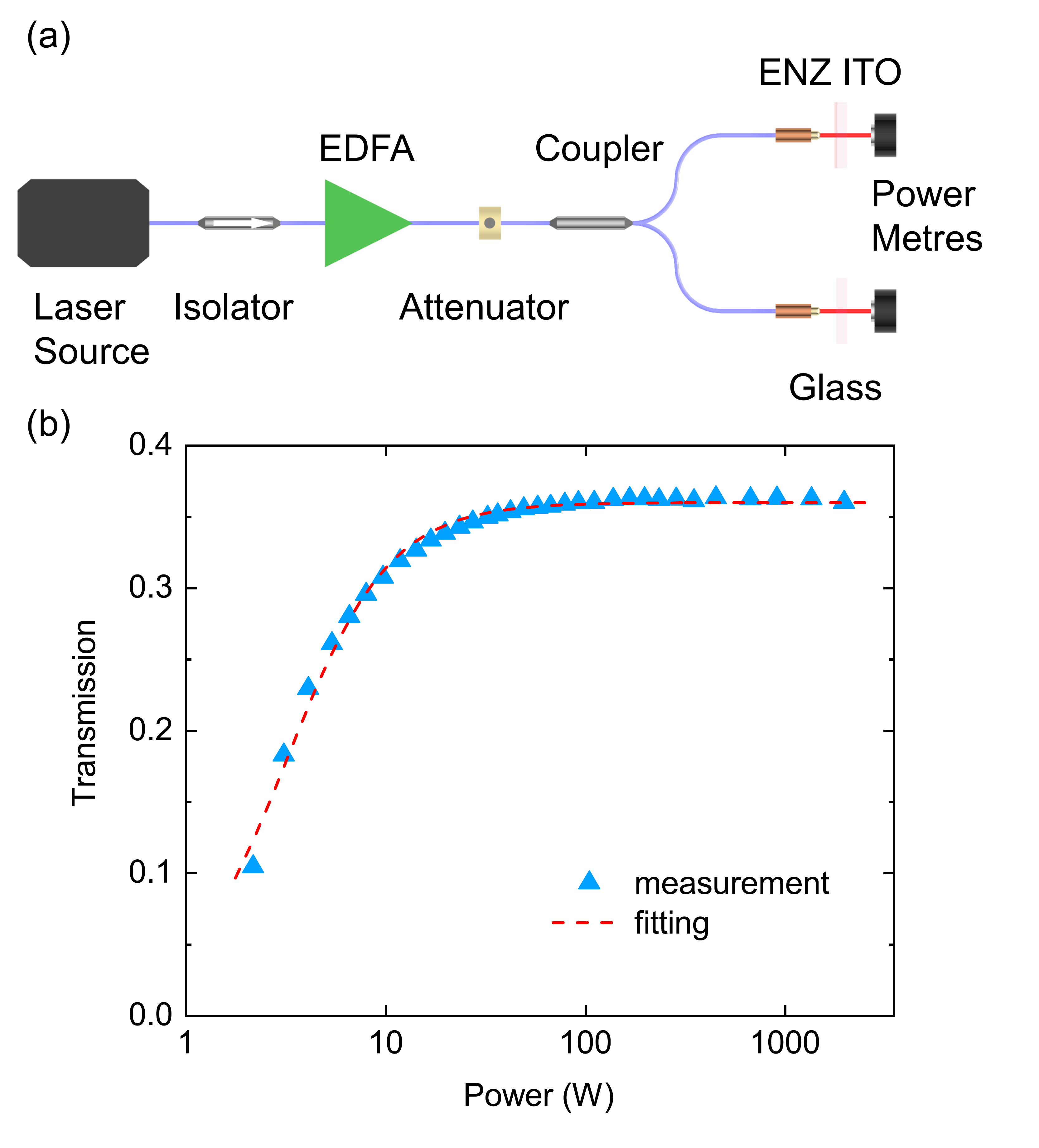} 
	\caption{. Nonlinear saturable absorption measurement. a) The experimental setup; b) The measured results. The triangles denote the data points, and the dashed line represent the fitting curve.}
	\label{fs3}
\end{figure}

The experimental setup and results of nonlinear saturable absorption measurement are shown in Fig. \ref{fs3}. The laser source used is a 1-ps 56-MHz repetition rate fiber laser at C-band, and the ratio of the coupler is 8.65 (ITO path) : 1 (glass path). The maximum absorption of 65\% ($A_{\rm sat} = 1-T_{\rm sat} = 1-0.35$) saturates at a peak power of $\sim$33.44 W. The data points are fitted using $T = T_{\rm sat}x^n/(k^n+x^n)$, with $k = 3.19905$ and $n = 1.68625$.

\section{The ML wavelength with ITO under different original ML wavelength}
\label{SI7}

\begin{figure}[h]
	\centering
	\includegraphics[width=0.5\linewidth]{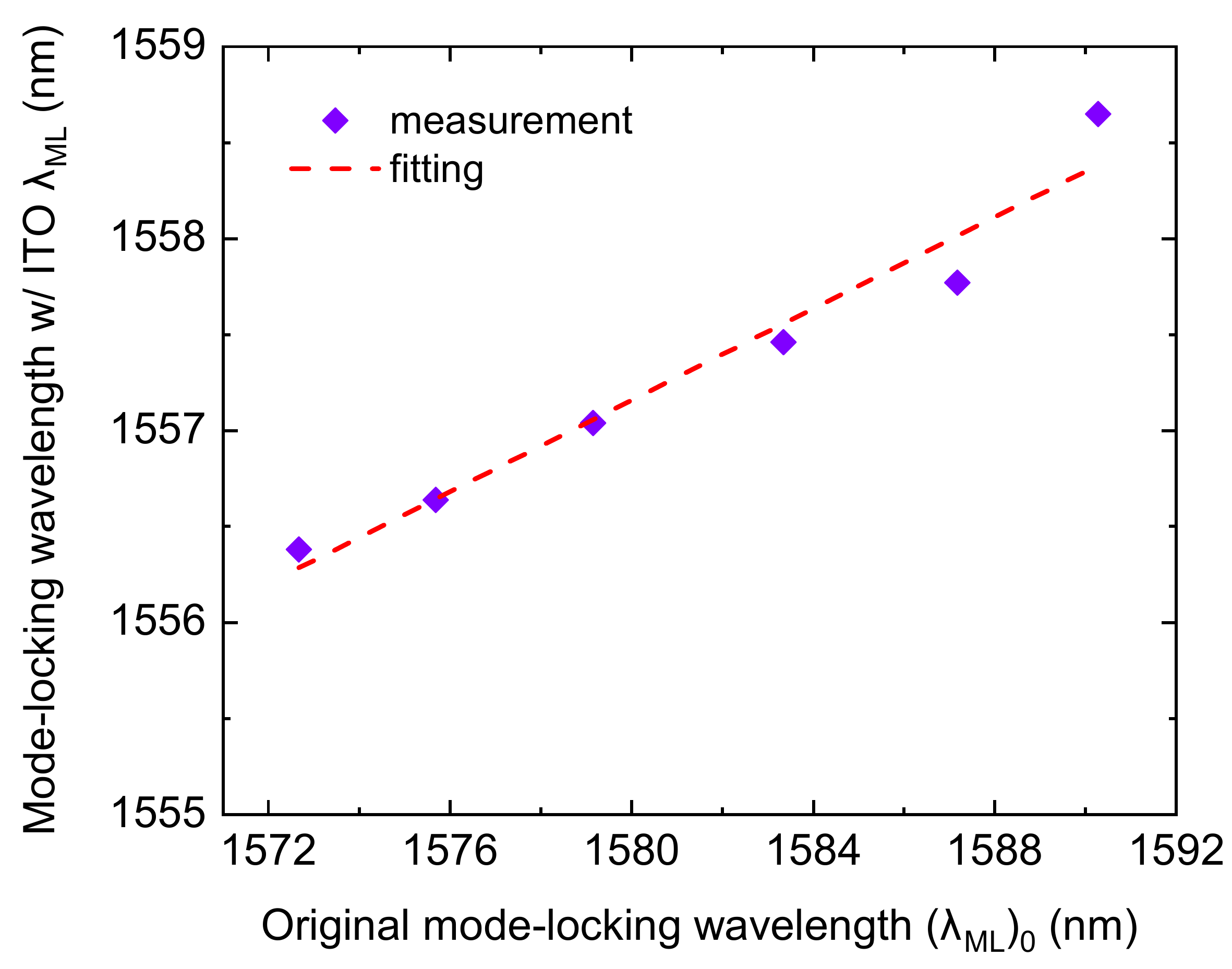} 
	\caption{. $\lambda_{\mathrm{ML}}$ versus $(\lambda_{\mathrm{ML}})_0$. The diamonds denote the data points, and the dashed line represent the fitting curve.}
	\label{fs4}
\end{figure}

In this Section, we measure the ML wavelength ($\lambda_{\mathrm{ML}}$) with ENZ ITO inserted ($\lambda_{\mathrm{ENZ}} = 1538$ nm) under different original mode-locking wavelengths ($(\lambda_{\mathrm{ML}})_0$). The results are shown in Fig. \ref{fs4}. One can see that ENZ ITO can stabilize $(\lambda_{\mathrm{ML}})_0$ to a smaller spectral region.

\section{The empirical formula}
\label{SI8}

The empirical formula in Eq. (3) in the main text takes the form:

\begin{equation}
	\mathcal{P}={a_{1}}\mathrm{Re}\left( {{n_{\mathrm{C}}}}\right) +{a_{2}}\mathrm{Im}\left( {{n_{\mathrm{C}}}}\right) +{a_{3}}{n_{\mathrm{f}}},
	\label{S2}
\end{equation}%

In its full form, by substituting the Drude parameters, it becomes 

\begin{equation}
	\begin{aligned} \displaystyle\mathcal{P} & = \displaystyle {a_1}\sqrt {\displaystyle \frac{{{\varepsilon _{\mathrm{b}}}}}{2} + \displaystyle \frac{{\sqrt {\left( {{\varepsilon _{\mathrm{b}}} - \displaystyle \frac{{\omega _{\mathrm{p}}^2}}{\Xi }} \right) + \Lambda } }}{2} - \frac{{\omega _{\mathrm{p}}^2}}{{2\Xi }}} \\ & + {a_2}\displaystyle \frac{{\gamma \omega _{\mathrm{p}}^2}}{{2\omega \Xi \sqrt {\displaystyle \frac{{{\varepsilon _{\mathrm{b}}}}}{2} - \displaystyle \frac{{\omega _{\mathrm{p}}^2}}{{2{\gamma ^2} + 2{\omega ^2}}} + \displaystyle \frac{{\sqrt {{{\left( {{\varepsilon _{\mathrm{b}}} - \displaystyle \frac{{\omega _{\mathrm{p}}^2}}{\Xi }} \right)}^2} + \Lambda }}}{2}} }} \\ & + {a_3}{n_{\mathrm{f}}}, \end{aligned}  
	\label{S3}
\end{equation}

\noindent where

\begin{equation}
	\begin{array}{l}
		\displaystyle \Lambda = \displaystyle \frac{{{\gamma ^2}\omega _{\mathrm{p}}^4}}{{{\omega ^2}{\Xi ^2}}}, \vspace{8pt}\\
		\displaystyle \Xi =\displaystyle {\gamma ^2} + {\omega ^2}.%
	\end{array}
	\label{S4}
\end{equation}

Equation (\ref{S2}) treats the height of the effective potential barrier as the weighted sum of all optical potentials, produced, in particular, by real and imaginary parts of the model's ingredients. Due to the fact that the longest optical path pertains to the optical fibers, the cavity's intrinsic optical potential $n_{\mathrm{f}}$ is set to 1.46. This value is taken as the refractive index of the single-mode fiber, the longest component, to represent the general refractive index of the cavity. The theoretical curve in Fig. 3(d) is calculated using Eq. (\ref{S3}), while the experimental curve is obtained by converting the refractive index, using Eq. (2) from the main text and plugging them in Eq. (\ref{S2}). The condition $\mathrm{d}\mathcal{P}/\mathrm{d}\lambda =0$ ensures the choice of the lowest effective potential barrier that we aim to find. 


\bibliography{References}